# The Portevin-Le Chatelier effect in the Continuous Time Random Walk framework

A. Sarkar* and P. Barat

Variable Energy Cyclotron Centre

1/AF Bidhan Nagar, Kolkata 700064, India

*We present a continuous time random walk model for the Portevin-Le Chatelier (PLC) effect. From our result it is shown that the dynamics of the PLC band can be explained in terms of the Levy Walk.*

The Portevin-Le Chatelier (PLC) effect is a type of plastic instability which manifests itself as stress serrations in the deformation characteristics of many dilute alloys [1-5]. This plastic instability results spatially inhomogeneous deformation, characterized by different types of strain localization which is generally called the PLC band [2-5]. It is widely accepted that this phenomenon stems from the metallurgical process, the dynamic strain aging (DSA) [6-8]. DSA is basically the manifestation of the dynamic interaction between the mobile dislocations and the diffusing solute atoms. The plastic deformation in these materials is primarily governed by the movement of the PLC band. During the plastic deformation, the gliding dislocation band is temporarily arrested at localized obstacles like forest dislocations and it overcomes the obstacles by the aid of the thermal activation. During the waiting time at the obstacles the band may undergo additional pinning by diffusing solute atoms [6-8]. The strength of the obstacle increases with the waiting time. The PLC effect is usually undesirable since it has detrimental influences like the loss of ductility and the appearance of surface markings on the specimen.

* Corresponding author: apu@veccal.ernet.in

The macroscopic spatio-temporal deformation dynamics that is associated with the PLC effect is quite complex. The complexity of the problem made the topic an interdisciplinary field of study. Several nonlinear dynamical and statistical studies have been carried out on the PLC effect over the last few years with the goal being to achieve a better understanding of the small-scale processes and of the multiscale mechanisms that link the mesoscale DSA to the macroscale PLC effect [9-11]. The characteristics of the dislocations causing plasticity in the PLC regime have also been the subject of many experimental studies using various techniques [12-15]. It is observed experimentally that after a certain strain, the PLC effect is solely governed by the movement of a single band [12]. To investigate the dynamics of the band we assume it is stationary at an obstacle for a random time interval $t$ prior to its release due to thermal activation. The band then performs a flight of a random length $x$ with a uniform velocity [13] before it gets trapped at another obstacle. This type of dynamics of the band movement can be described in terms of a continuous time random walk (CTRW) model. It can be fully specified by the probability distribution of waiting time $\psi(\mathrm{t})$ and flight lengths $\lambda(x)$. Thus, $\psi(t)dt$ is the probability that the band is trapped at a given position at time between $t$ and $t+dt$, and $\lambda(x)dx$ gives the probability that the band flights a distance between $x$ and $x+dx$ during an avalanche.

The CTRW, first introduced by Montroll and Weiss in 1965 [16], has a large number of applications to the modeling of many physical phenomena [17-19]. In contrast to the standard random walk in which steps are made periodically, the CTRW is based on the idea that the length of a given jump, as well as the waiting time elapsed between two successive jumps are regulated by a jump probability density function (pdf) $w(x,t)$. The function $w(x,t)$ determines the jump length and the waiting time pdfs as

$$\left.\begin{aligned} \lambda(x) &= \int_0^\infty w(x,t)dt \\ \psi(t) &= \int_0^\infty w(x,t)dx \end{aligned}\right\} \qquad (1)$$

CTRW processes are categorized by the characteristic waiting time

$$T = \int_0^\infty t\psi(t)dt \qquad (2)$$

and the jump length variance

$$\Sigma^2 = \int_0^\infty x^2 \lambda(x)dx \qquad (3)$$

being finite or divergent respectively. With these definitions, a CTRW process can be described by the equation

$$\eta(x,t) = \int_0^\infty dx' \int_0^t \eta(x',t')w(x-x',t-t')dt' + \delta(x)\delta(t) \qquad (4)$$

where $\eta(x,t)$ is the pdf of just having arrived at the position $x$ at time $t$ and $\delta(x)\delta(t)$ denotes the initial condition of the random walk. Consequently, the pdf $p(x,t)$ of being at $x$ at time $t$ is given by

$$p(x,t) = \int_0^\infty \eta(x,t')\phi(t-t')dt' \qquad (5)$$

where $\phi(t)$ is defined by the cumulative probability of having no jump during the time interval (0,t) and is given by,

$$\phi(t) = 1 - \int_0^t \psi(t')dt' \qquad (6)$$

In Fourier-Laplace space, the pdf $p(x,t)$ obeys the algebraic relation [20]

$$p(k,s) = \frac{1-\psi(s)}{s} \frac{p_0(k)}{1-w(k,s)} \qquad (7)$$

where $p_0(k)$ denotes the Fourier transform of the initial condition. This is called the Montroll-Weiss formula. To obtain, the pdf $p(x,t)$ we have to invert the Fourier Laplace transform in the limit $(k,s) \to (0,0)$, i.e.

$$p(x,t) = L^{-1}\left\{\lim_{s\to 0} F^{-1}\left\{\lim_{k\to 0} p(k,s)\right\}\right\} \qquad (8)$$

To understand the nature of the band dynamics in the frame work of CTRW we have carried out tensile tests on flat Al-2.5%Mg alloy at room temperature at various strain rates. The details of the experiment can be found in the Ref. [21]. Fig. 1 shows a typical segment of the stress-time curve for the strain rate $1.7\times10^{-3}$ $sec^{-1}$. The serrations observed in the stress-time curve are the outcome of the jerky propagation of the PLC band through the sample. The waiting time of the band corresponds to the time required for stress to increase and is designated as $t_w$ in the Fig. 1. Thus the waiting time during each step of the band movement can be easily obtained from the stress-time data and hence the waiting time distribution. On the other hand, the jump length of the band is related to the strain carried by it. The exact estimation of the strain at each band jump is very difficult. Several experiments have been carried out using sophisticated techniques to determine various band parameters [12-15]. However, the precise measurement of the strain carried by the PLC band at each jump is still lacking.

We have calculated the waiting time distribution from the experimental stress-time data. A typical plot of the waiting time distribution for the strain rate $1.7\times10^{-3}$ $sec^{-1}$ is shown in Fig. 2. The waiting time distribution is found to show a power law behavior.

Distributions $\psi(t)$ of events with property $t$ that behave like $\psi(t)\sim t^{-\gamma}$ , as observed in our experimental data of waiting time, are most conveniently analyzed with the help of the integrated (cumulative) distribution

$$\overline{W}(t) = \int_{t}^{M} \psi(t)dt \qquad (9)$$

where $M$ is the maximal event encountered in the data set. By using the integrated description instead of histograms we avoid data fluctuations in the high value regime induced by the choice of linear bins. If $M\longrightarrow\infty$ (and if $\gamma>1$) then $\overline{W}(t) \sim t^{-\gamma+1}$. The waiting time obtained from the experimental data is generally confined to the ranges $0.05<t<3$ seconds. Therefore, we cannot replace M by $\infty$ in Eqn. (9) and obtain

$$W(t) = \overline{W}(t)/t \sim \frac{1}{t^{\gamma}}\left[1-\left(\frac{t}{M}\right)^{(\gamma-1)}\right] \qquad (10)$$

Thus, the log-log plot of $W(t)$ vs. $t$ definitely departs from a straight line as $t$ approaches $M$.

All the cumulative distributions of the waiting time obtained from the experimental data fitted extremely well with the Eqn. (10). A typical fit for the strain rate $1.7\times10^{-3}$ $sec^{-1}$ is shown in the inset of the Fig. 2. From the fit we have obtained the power law exponent $\gamma$. The values of the $\gamma$ for the different strain rates are found to lie between 2 and 3. So in the asymptotic limit we can represent the waiting time distribution as

$$\psi(t) \sim \left(\frac{T}{t}\right)^{\gamma} \qquad (11)$$

In this limit of $\gamma$ the charateristic waiting time (T) is finite but the second moment of the waiting time diverges. The distribution with this asymptotic form is in general referred to as the Levy distribution.

Fig. 3 shows a typical scatter plot of stress drop magnitude against the waiting time for the experiment conducted at the strain rate $1.7\times10^{-3}$ $\sec^{-1}$. The plot clearly reveals a positive correlation between the waiting time and the stress drop magnitude. This suggests that large waiting time increases the probability of larger stress drop to occur. Again the jump length of the PLC band is one of the important parameter to decide the magnitude of the stress drop. Thus, it is straight forward to assume that the jump length and waiting time distributions are coupled. To describe the movement of the band in the CTRW framework we choose jump probability density function as

$$w(x,t) = \lambda(x\,|\,t)\psi(t) \qquad (12)$$

where $\lambda(x\,|\,t)$ is the conditional probability density of making a jump of length $x$ given that the waiting time is $t$. Following the idea of Zumofen et. al. [22], to account for the constant velocity of the band movement, we choose

$$\lambda(x\,|\,t) \sim \delta(|\,x\,| - t) \qquad (13)$$

Thus,

$$w(x,t) \sim \delta(|\,x\,| - t)\psi(t) \qquad (14)$$

$\psi(t)$ is given by the Eqn. (11). Eqn. (14) allows jump of arbitrary length, but long jumps are penalized by requiring more time to be performed. These types of processes with waiting time distribution having divergent higher order moments and coupled jump pdf are designated as Levy-walk [22-24].

Combining Eqns. (7), (11) and (14), allows one to calculate $p(k,s)$ and finally $p(x,t)$ from Eqn. (8). Once $p(x,t)$ is known, in principle all statistical quantities can be calculated. In this case, the mean-squared displacement has been shown to follow [22,23]

$$\left\langle x^2(t) \right\rangle \sim \begin{cases} t^{4-\lambda}.2<\gamma<3 \\ t\ln t, \gamma=3 \\ t, \gamma>3 \end{cases} \qquad (15)$$

Thus, in case of the PLC band dynamics,

$$\left\langle x^2(t) \right\rangle \sim t^{4-\gamma} \qquad (16)$$

This establishes the fact that the PLC band exhibits Levy walk type of anomalous enhanced diffusion.

In a recent paper [21], we have carried out detailed scaling analysis of the PLC effect using two complementary scaling analysis methods, diffusion entropy analysis and variance method [25,26]. From the results of the scaling analysis, we could establish that the PLC effect exhibit persistent Levy walk type of scaling behavior. Here we show that if we put the PLC effect in the CTRW framework with space-time coupled jump pdf, the Levy walk property is revealed in the PLC band dynamics.

**Acknowledgement**

We gratefully acknowledge the fruitful discussions with Dr. Nicola Scafetta.

**Figure Captions**

Fig. 1. A typical portion of the stress-time curve of the Al-2.5%Mg alloy deformed at a strain rate $1.7\times10^{-3}$ $sec^{-1}$. The inset shows a small segment in it.

Fig. 2. The typical waiting time distribution plot (log-log scale) for the strain rate $1.7\times10^{-3}$ $sec^{-1}$. The inset shows the cumulative distribution plot fitted with the Eqn. (10).

Fig. 3. The typical scatter plot of stress drop magnitude against waiting time for the strain rate $1.7\times10^{-3}$ $sec^{-1}$.

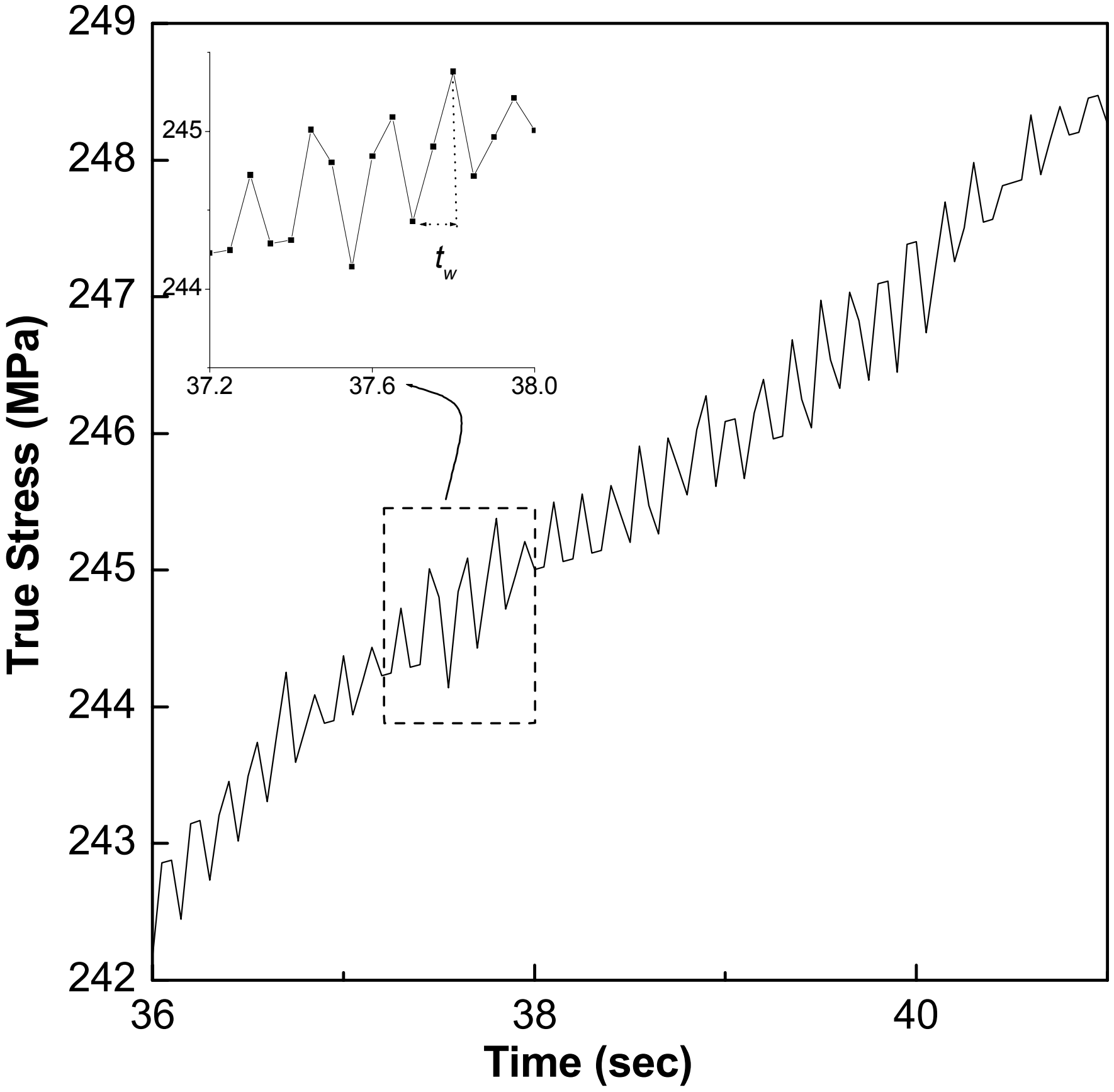

Fig. 1

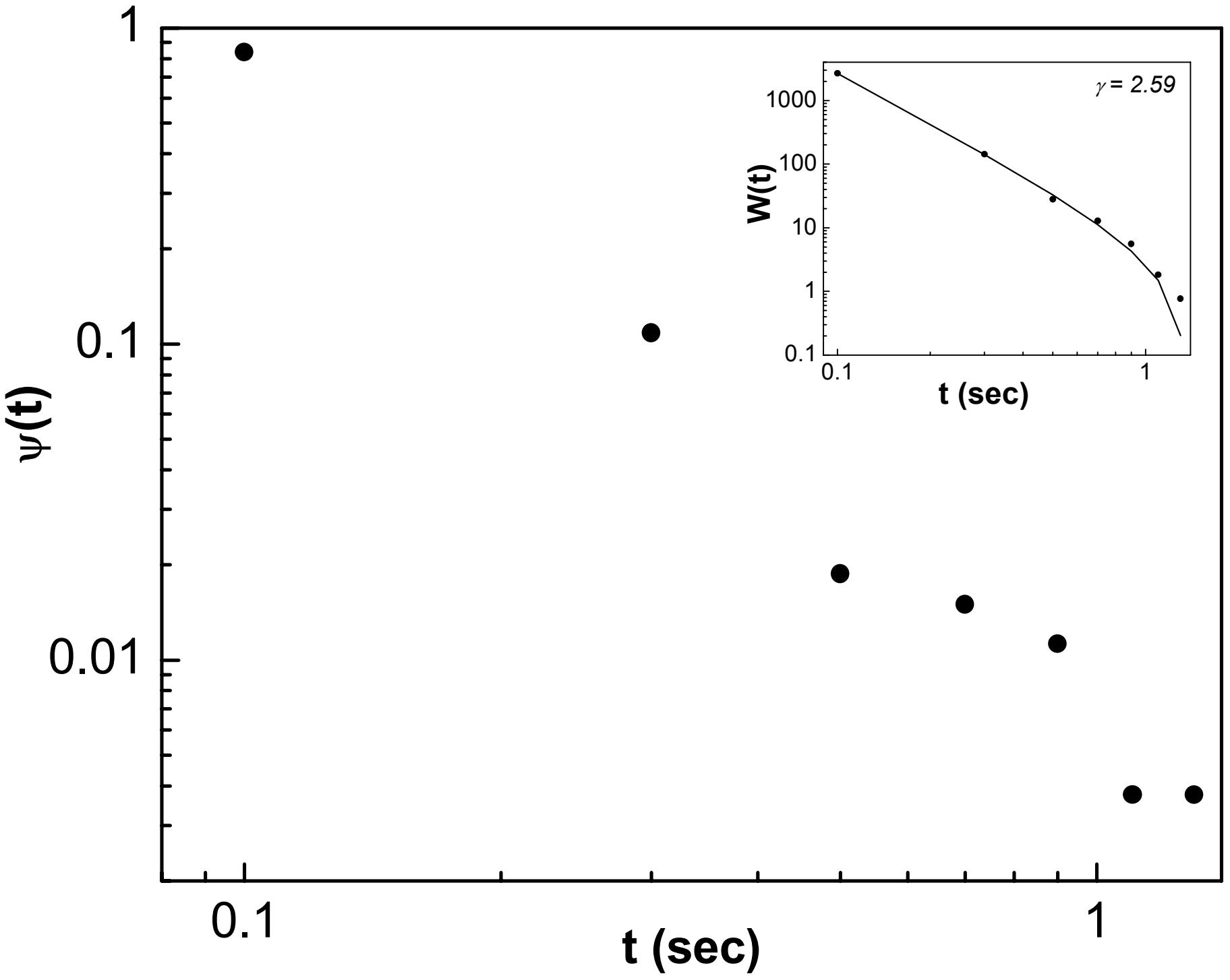

Fig. 2

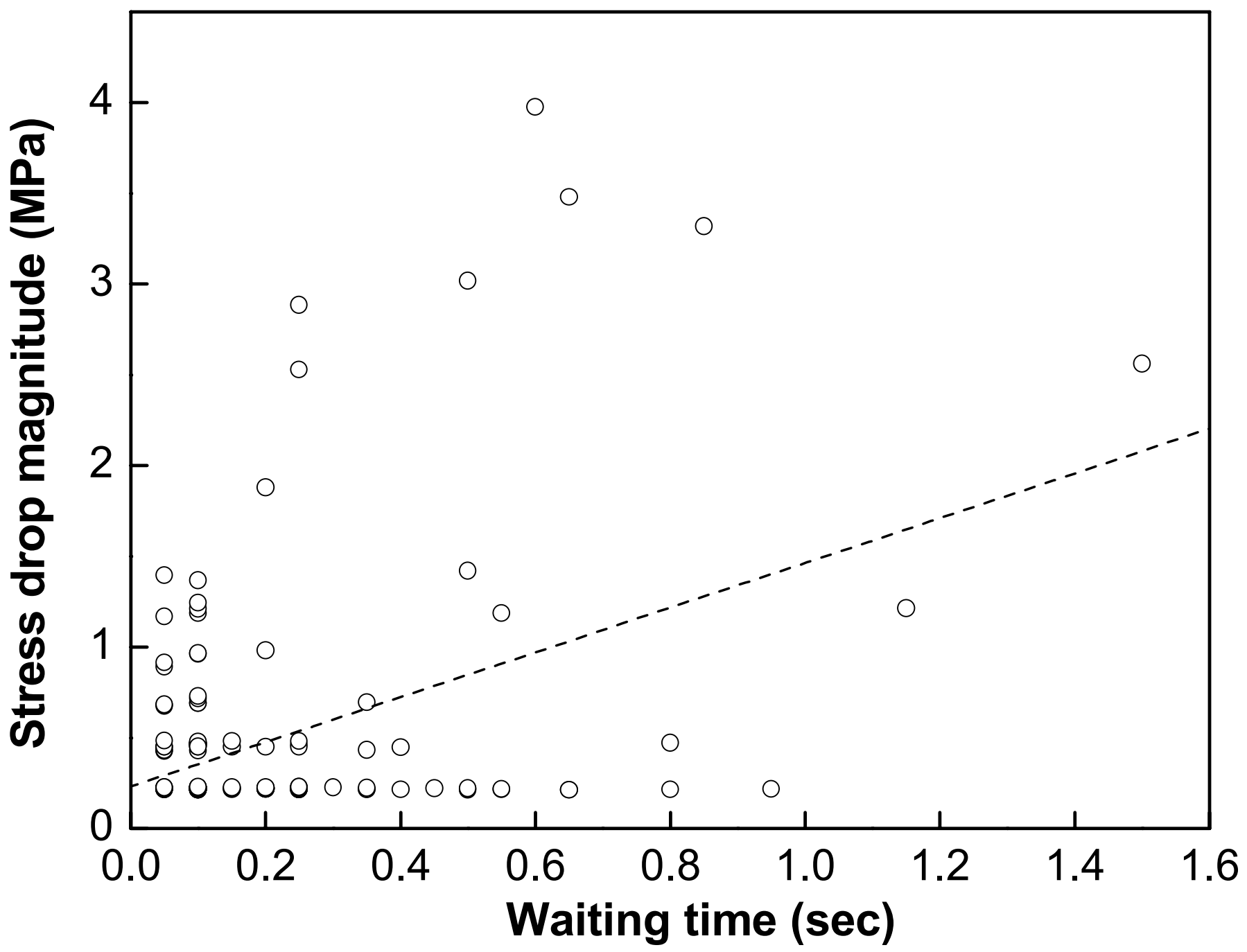

Fig. 3